\documentclass[reprint,twocolumn,superscriptaddress,showpacs,floatfix]{revtex4-1}
\usepackage{amsmath}
\usepackage{mathrsfs}
\usepackage{txfonts}
\usepackage{amssymb}
\usepackage{graphicx}
\usepackage{hyperref}
\usepackage{ulem}
\usepackage{overpic}
\usepackage{psfrag}
\usepackage{tabularx}
\usepackage{multirow}
\usepackage{array}
\usepackage{placeins}
\usepackage[marginal]{footmisc}

\newcommand{\PreserveBackslash}[1]{\let\temp=\\#1\let\\=\temp}

\makeatletter

\begin{document}
\title{An alternative spontaneous symmetry breaking pattern for $\rm{U}(1)$ with no gapless Goldstone mode}

\author{Huan-Qiang Zhou}
\affiliation{Centre for Modern Physics, Chongqing University, Chongqing 400044, The People's Republic of China}
\author{Qian-Qian Shi}
\affiliation{Centre for Modern Physics, Chongqing University, Chongqing 400044, The People's Republic of China}
\author{Yan-Wei Dai}
\affiliation{Centre for Modern Physics, Chongqing University, Chongqing 400044, The People's Republic of China}

\begin{abstract}
An emergent gapless Goldstone mode originates from continuous spontaneous symmetry breaking, which has become a doctrine since the pioneering work by Goldstone [J. Goldstone, Nuovo Cimento \textbf{19}, 154 (1961)]. However,  we argue that it is possible for a continuous symmetry group  $\rm{U}(1)$ to make an exceptional case, simply due to the well-known mathematical result that a continuous symmetry group $\rm{U}(1)$ may be regarded as a limit of a discrete symmetry group $Z_q$ when $q$ tends to infinity. As a consequence, 
spontaneous symmetry breaking for such a continuous symmetry group $\rm{U}(1)$ does not necessarily lead to any gapless Goldstone mode. This is explicitly explained for an anisotropic extension of the ferromagnetic spin-1 biquadratic model. In a sense, this model provides an illustrative example regarding the dichotomy between continuity and discreteness.

\end{abstract}
\maketitle

\section{Introduction}

Distinct types of quantum states of matter emerge from spontaneous symmetry breaking (SSB) - a fundamental notion in the conventional Landau-Ginzburg-Wilson paradigm~\cite{SSB}. In particular, if a continuous symmetry group is spontaneously
broken, then an emergent gapless Goldstone mode (GM)~\cite{GM} appears in an attempt to recover the broken symmetry. This is in sharp contrast to SSB for a discrete symmetry group. In a sense, this results in a dichotomy between continuous symmetry groups and discrete symmetry groups, as far as SSB is concerned.

For a relativistic system undergoing SSB,
the number of  GMs is equal to the number of broken symmetry generators $N_{BG}$. However, for a non-relativistic system,  the connection between the number of GMs and the number of broken symmetry generators $N_{BG}$ is much more involved~\cite{Chadha,schafer,miransky,nicolis,brauner-watanabe,watanabe,watanabe2,Brauner,NG,NG1,NG2}. A proper classification of GMs requires 
to introduce type-A and type-B GMs~\cite{watanabe,watanabe2}, as a further development of a previous observation made by Nambu~\cite{nambu}. As a consequence,
one is led to the counting
rule of GMs that $N_A+2N_B=N_{BG}$, when the symmetry group $G$ is spontaneously broken into $H$, where $N_A$ and $N_B$ are, respectively, the numbers of type-A
and type-B GMs, and $N_{BG}$ is equal to the dimension of the coset space $G/H$. In a sense, this classification partially resolves a long-standing debate between Anderson and Peierls
concerning whether or not the ${\rm SU}(2)$ ferromagnetic states originate from SSB~\cite{Anderson}, given that the ${\rm SU}(2)$ ferromagnetic states constitute a prototype
for SSB from ${\rm SU}(2)$ to ${\rm U}(1)$, with the emergence of one type-B GM.

A natural question arises as to whether or not the dichotomy between continuity and discreteness exhausts all the possible types of SSB that occurs, in principle, in quantum many-body systems. Actually, in our opinion, SSB for a symmetry group ${\rm U}(1)$ occupies a prominent position, since it {\it only} leads to a type-A GM in two- and higher dimensional quantum many-body systems, if one takes both the Mermin-Wagner-Coleman theorem~\cite{mwc} and the counting rule of GMs~\cite{watanabe,watanabe2} into account. As any other patterns for SSB with type-A GMs,
it {\it only} occurs in the thermodynamic limit, in contrast to SSB with type-B GMs. In fact, this SSB pattern is visualized as the Mexican hat in the energy configuration. 

Conventionally, SSB is characterized in terms of a singularity arising from the two non-commutative limiting operations~\cite{berry}: one is the thermodynamic limit, and the other is to demand that an additional term in the model Hamiltonian, with its density acting as a (local) order parameter and explicitly breaking the symmetry group $G$, vanishes. However, if $G= {\rm U}(1)$, the situation becomes quite subtle, due to the fact that ${\rm U}(1)$ may be regarded as a limit of a cyclic group ${\rm Z}_q$,  when $q$ tends to infinity. Hence, a SSB pattern for ${\rm U}(1)$ arises from a singularity due to the non-commutativity of the two limiting operations: one is $L \rightarrow \infty$, and the other is $q \rightarrow \infty$. In fact, such a SSB pattern for ${\rm U}(1)$ occurs in a superfluid phase, with one type-A GM, in two- and higher dimensions. However, one may imagine an alternative SSB pattern for a continuous symmetry group ${\rm U}(1)$, if $q$ is simply related with the system size $L$. 
In other words, the two limiting operations, i.e., $L \rightarrow \infty$ and $q \rightarrow \infty$, are essentially identical.
If so, an alternative SSB pattern for a continuous symmetry group $\rm{U}(1)$ occurs, with a salient feature that no gapless GM emerges.  

In this work, we demonstrate that such an alternative SSB pattern for a continuous symmetry grop $\rm{U}(1)$ does occur in quantum many-body systems, thus leading to an exotic quantum state of matter. As it turns out, the coexisting fractal (CF) phases in the spin-1 ferromagnetic anisotropic biquadratic model offer an illustrative example for this pattern, thus leading to
infinitely degenerate (unentangled) factorized ground states that are scale-invariant in the thermodynamic limit, with the fractal dimension $d_f$, identical to the number of type-B GMs $N_B$, being zero.

\section{The ground state phase diagram for the spin-1 ferromagnetic anisotropic biquadratic model}

\begin{figure}
     \includegraphics[angle=0,totalheight=5cm]{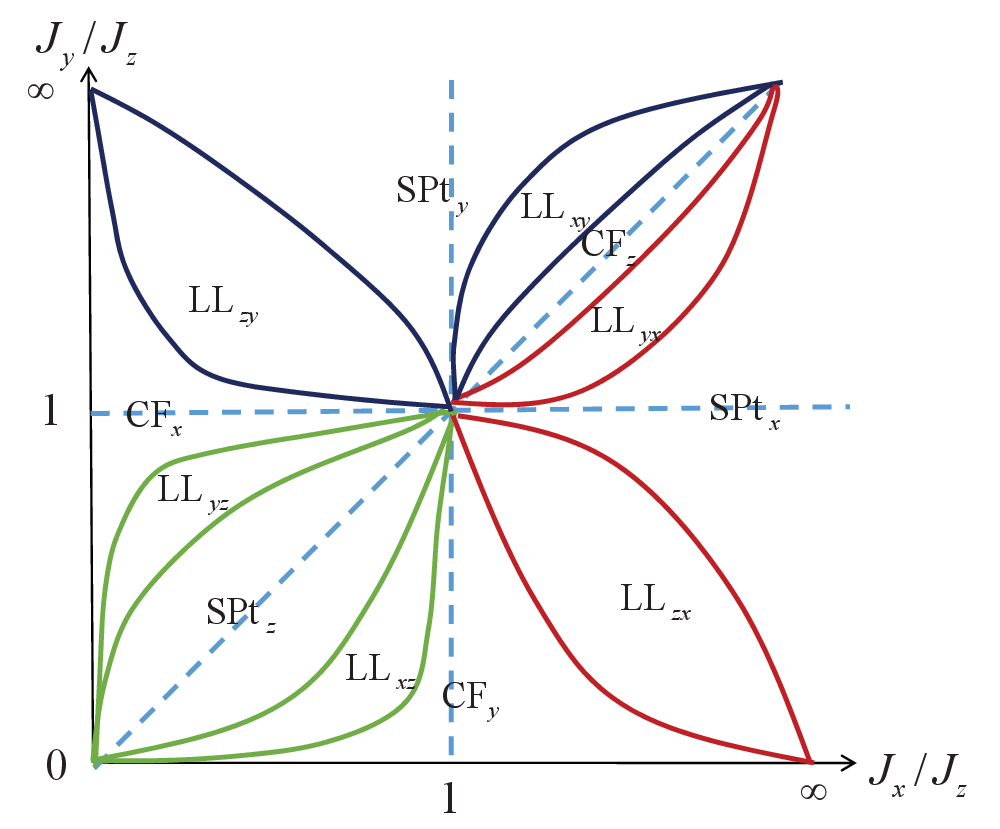}
       \caption{ A sketch of the ground state phase diagram for an anisotropic extension of the spin-1 ferromagnetic biquadratic model (\ref{xyz2}), which is adapted from
       Ref.~\cite{spin1BM} (also cf.~Ref.\cite{entropy}). We focus on the region: $J_x/J_z \geq 0$ and $J_y/J_z \geq 0$, due
       to a  symmetric consideration. Here, a solid line indicates a phase transition line.
There are twelve distinct phases: three CF phases labeled as $\rm{CF_{x}}$, $\rm{CF_{y}}$ and $\rm{CF_{z}}$, six LL phases
labeled as $\rm{LL_{xy}}$, $\rm{LL_{yz}}$, $\rm{LL_{zx}}$, $\rm{LL_{yx}}$, $\rm{LL_{xz}}$ and $\rm{LL_{zy}}$, and three SPt phases labeled as
$\rm{SPt_{x}}$, $\rm{SPt_{y}}$  and $\rm{SPt_{z}}$, respectively.
Note that both horizontal and vertical axes are in a scale, defined by $\arctan(J_x/J_z)$ and $\arctan(J_y/J_z)$, respectively.
}\label{phasediagramfxyz2}
\end{figure}

An anisotropic extension of the spin-1 ferromagnetic biquadratic model~\cite{bqm} is described by the Hamiltonian
\begin{equation}
\mathscr{H}=\sum_{j}{(J_xS_j^x S_{j+1}^x+J_yS_j^yS_{j+1}^y+J_zS_j^zS_{j+1}^z)^2}, \label{xyz2}
\end{equation}
where $S_j^x$, $S_j^y$, and $S_j^z$  are the spin-1 operators at a lattice site $j$, and $J_x$, $J_y$, and $J_z$ denote the coupling parameters describing anisotropic interactions.
The sum over $j$ is taken from 1 to $L-1$ under the open boundary conditions (OBCs) and from 1 to $L$ under the periodic boundary conditions (PBCs). Its isotropic version is an exactly solvable case in the spin-1 bilinear-biquadratic model~\cite{blbq}. 
At a generic point in the parameter space, the model Hamiltonian (\ref{xyz2}) possesses the symmetry group ${\rm U}(1) \times {\rm U}(1)$, generated by $K_{yz} \equiv \sum_j K_{yz}^j$ and $K_{x} \equiv \sum_j K_{x}^j$, with $K_{yz}^j= \sum _j (-1)^{j+1} [(S_j^{y})^2-(S_j^{z})^2]$ and $K_{x}^j=\sum _j (-1)^{j+1} (S_j^{x})^2$, respectively.  However, on the three characteristic lines
($J_x=J_y, J_y= J_z$ and $J_z=J_x$), the symmetry group is enlarged to ${\rm SU}(2) \times {\rm U}(1)$~\cite{spin1BM} (for more details, we refer to Sec. A of the Supplemental Material (SM)). 
The symmetry generators are staggered, thus we have to restrict ourselves to even $L$'s.
The ground state phase diagram is sketched in Fig.~\ref{phasediagramfxyz2}, which is adapted from Ref.~\cite{spin1BM} (also cf.~Ref.~\cite{entropy}), as a result of the numerical simulation
in terms of the iTEBD algorithm~\cite{TN2}. There are twelve distinct phases: three CF phases, labeled as $\rm{CF_{x}}$, $\rm{CF_{y}}$ and $\rm{CF_{z}}$, six
Luttinger liquid (LL) phases, with central charge $c=1$, labeled as $\rm{LL_{xy}}$, $\rm{LL_{yz}}$, $\rm{LL_{zx}}$, $\rm{LL_{yx}}$, $\rm{LL_{xz}}$ and $\rm{LL_{zy}}$,
and three symmetry-protected trivial (SPt) phases~\cite{spt}, labeled as $\rm{SPt_{x}}$, $\rm{SPt_{y}}$  and $\rm{SPt_{z}}$, respectively.
As demonstrated in Ref.~\cite{spin1BM}, a novel universality class arises from instabilities of the LL  phases towards the CF phases.  In addition, QPTs between the
LL phases and the SPt phases are identified to be in the KT universality class.

Before proceeding, we remark that the three CF phases, six LL phases and three SPt phases are dual to each other under duality transformations induced from the permutation group $S_3$ with respect to $S_j^x, S_j^y$ and $S_j^z$: $S_j^x\leftrightarrow S_j^y$,  $S_j^y\leftrightarrow S_j^z$ and $S_j^x\leftrightarrow S_j^z$~\cite{spin1BM}. Hence, we restrict ourselves to characterize the $\rm{CF_{x}}$ phase, the $\rm{LL_{yz}}$ phase, and the $\rm{SPt_{z}}$ phase. 

We start with the two characteristic lines $J_y/J_z=1$ and $J_x=0$, located inside the $\rm{CF_{x}}$ phase.
On the characteristic line $J_y/J_z=1$, highly degenerate ground states occur, with the ground state energy per lattice site being equal to $J_x^2$ and the ground state degeneracy being $L+1$,  as a result of SSB from $\rm{SU(2)\times \rm{U(1)}}$ to $\rm{U(1)\times
\rm{U(1)}}$~\cite{shiqq1,FMGM, spin1BM}, with one type-B GM~\cite{spin1BM}. 
Hence, the highly degenerate ground states are scale-invariant, characterized in terms of the fractal dimension $d_f$~\cite{doyon} (also cf. Ref.~\cite{popkov}). 
As argued in Refs.~\cite{FMGM, spin1BM}, the entanglement entropy exhibits a logarithmic scaling relation with the block size $N$ in the thermodynamic limit $L \rightarrow \infty$, with the prefactor being half the number of type-B GMs $N_B$, thus leading to the identification of the fractal dimension $d_f$ with the number of type-B GMs $N_B$.  We remark that both the highest and lowest weight states are (unentangled) factorized states,  with the entanglement entropy being zero.

Now we turn to the characteristic line $J_x=0$ with $J_y/J_z>0$ in the $\rm{CF_{x}}$ phase.
On this characteristic line, there are a two-parameter family of factorized  ground states $|\phi_f\rangle$,  with
the ground state degeneracy being $L+1$, and the ground state energy per lattice site being equal to $0$.
The explicit expression for $|\phi_f\rangle$ has been presented in Ref.~~\cite{spin1BM} (also cf. Sec. A of the SM). Instead,
we are interested in a particular factorized ground state
$|\phi_0\rangle=\bigotimes_j |v\rangle_j$, where $|v \rangle_j=\mu |0_y\rangle_j + \nu |0_z\rangle_j$, with $\mu^2+\nu^2=1$ and  $\mu=\sqrt {J_y/(J_y+J_z)}$. Here, $|0_y\rangle_{j}$ and $|0_z\rangle_{j}$ are basis states, with an eigenvalue being zero, for the spin operators $S^y_{j}$ and $S^z_{j}$, respectively.
Actually,  a two-parameter family of factorized  ground states $|\phi_f\rangle$ are generated from the action of the symmetry group 
${\rm U}(1) \times {\rm U}(1)$ on $|\phi_0\rangle$. Note that
the ground state  $|\phi_0\rangle$ is invariant under the one-site translation operation if PBCs are adopted or
under the permutation $P_{12}P_{23}\cdots,P_{L-1,L}$ if OBCs are adopted, where
$P_{kk+1}$ ($k=1,\cdots,L-1$) denote the generators of the permutation group $S_L$. 
Moreover, $|\phi_0\rangle$ evolves into the highest weight state $|\otimes_{j=1}^L\{1_x\}_j\rangle$ for the symmetry group $\rm{SU}(2)$, as $J_y$ tends to $J_z$.
Here, $|1_x\rangle$ denotes the eigenvector of $S_j^x$, with the eigenvalues being $1$.

It is convenient to introduce $q$ H-orthogonal states  $|\psi_k\rangle$ ($k=0,1,\cdots,L$), defined as $|\psi_k\rangle \equiv (V_q)^k |\phi_0\rangle$, where $V_q$ denotes an operator $V_q=\prod _j\exp(i2\pi K_{yz}^j/q)$, with $q=L+1$. Indeed, $V_q$ itself is an element of a cyclic group ${\rm Z}_{q}$, which turns out to be a subgroup of the symmetry group $\rm{U}(1)$ generated by $K_{yz}$.
Here, we mention that the notion of $q$-orthogonal states has been introduced  to describe SSB for discrete symmetry groups in Ref.~\cite{shiqq2}.  Hence, it is the cyclic group ${\rm Z}_{q}$ that connects the $q$-orthogonal states  $|\psi_k\rangle$, which becomes ${\rm U}(1)$ in the thermodynamic limit.
As a result, an alternative SSB pattern for a continuous symmetry group ${\rm U}(1)$ occurs on the characteristic line $J_x=0$ with $J_y/J_z>0$, with a salient  feature that no gapless GM emerges. 

A remarkable fact is that the highly degenerate ground states are permutation-invariant with respect to the unit cells consisting of the two nearest-neighbor lattice sites on the two characteristic lines $J_y/J_z=1$ and $J_x=0$.  That is, there is an {\it emergent} permutation symmetry group $S_{L/2}$ in the ground state subspace, given the Hamiltonian~\ref{xyz2} is not permutation-invariant. In fact,  the $\rm{CF_{x}}$ phases may be attributed to the coexistence of ${\rm SU}(2)$ SSB with one type-B GM~\cite{brauner-watanabe, watanabe, watanabe2} on the characteristic line  $J_y/J_z=1$ and an alternative SSB pattern for ${\rm U}(1)$ without any gapless GM on the characteristic line  $J_x=0$ with $J_y/J_z>0$.

\section{The ground state and the low-lying states}

We aim to reveal distinct features for the three phases - the $\rm{CF_{x}}$ phase, the $\rm{LL_{yz}}$ phase, and the $\rm{SPt_{z}}$ phase, by performing numerical simulations in terms of
the exact diagonalization (ED). For a chosen value of the system size $L$, we focus on the ground state $|\varphi_0\rangle$ and (up to $L$) low-lying states, labeled as  $|\varphi_k\rangle$, with $k=1,\cdots, L$.
 
In the $\rm{SPt_{z}}$ phase, we evaluate the ground state $|\varphi_0\rangle$ and a few (up to $L$) low-lying states $|\varphi_k\rangle$, together with their eigenvalues $E_k$ ($k=0,1,\cdots, L$).  
There is only a unique ground state, and the other low-lying states do not exhibit any pattern (for more details, cf. Sec. B of the SM). Further, we evaluate the  entanglement entropy $S(L,N)$, as a function of the block size $N$, for the point $(J_x/J_z, J_y/J_z)=(0.2, 0.3)$  in the $\rm{SPt_{z}}$ phase. 
It is found that the entanglement entropy $S(L,N)$  saturates, as the block size $n$ increases.  Thus, the $\rm{SPt_{z}}$ phase is gapful (for more details, cf. Sec. C of the SM).

In the $\rm{LL_{yz}}$ phase,  we evaluate the ground state $|\varphi_0\rangle$ and $L$ low-lying states $|\varphi_k\rangle$ ($k=1,\cdots, L$) for a few chosen values of $L$,
together with their eigenvalues $E_k$ ($k=0,1,\cdots, L$).    
It is found that the $l$-th and  $L+1-l$-th low-lying states, i.e.,  $|\varphi_{l}\rangle$ and  $|\varphi_{L+1-l}\rangle$, are degenerate in pair, 
 with the energy eigenvalues $E_l = E_{L+1-l}$ ($l=1,\cdots, L/2$),  where $l$ ranges from 1 to $L/2$,
in addition to the non-degenerate ground state  $|\varphi_0\rangle$.
The energy gap $\Delta_{0L/2}(L)$ between the ground state $|\varphi_0\rangle$ 
and the $L/2$-th low-lying state $|\varphi_{L/2}\rangle$, denoted as $\Delta_{0L/2}(L)$, scales as $\Delta_{0L/2}(L)\sim L$,
thus indicating that the ratio  $\Delta_{0L/2}(L)/L$ survives, as $L$ increases, in the $\rm{LL_{yz}}$ phase (for more details, cf. Sec. B of the SM).
In order to lend further support for the fact that the $\rm{LL_{yz}}$ phase is a LL phase, the finite-size matrix product state (MPS) algorithm~\cite{DM3} has been 
exploited to extract central charge $c$. In fact, a finite-size scaling analysis of the entanglement entropy yields $c = 1$~\cite{spin1BM} (for more details, cf. Sec. C of the SM).

In the ${\rm CF}_x$ phase, we evaluate the ground state $|\varphi_0\rangle$ and $L$ low-lying states $|\varphi_k\rangle$ ($k=1,\cdots, L$) for a few chosen values of $L$,
together with their eigenvalues $E_k$ ($k=0,1,\cdots, L$).   
It is found that  the $l$-th and  $L+1-l$-th low-lying states, i.e.,  $|\varphi_{l}\rangle$ and  $|\varphi_{L+1-l}\rangle$, are degenerate in pair, with the energy eigenvalues $E_l = E_{L+1-l}$ ($l=1,\cdots, L/2$),  where $l$ ranges from 1 to $L/2$, in addition to 
the non-degenerate ground state  $|\varphi_0\rangle$.  This pattern is exactly identical to that in  the $\rm{LL_{yz}}$ phase. However, the energy gap between the ground state $|\varphi_0\rangle$  and the $L/2$-th low-lying state $|\varphi_1\rangle$, denoted as  
$\Delta_{0L/2}(L)$, scales in a drastically different way, and the ratio  $\Delta_{0L/2}(L)/L$ vanishes in the thermodynamic limit $L \rightarrow \infty$ (for more details, cf. Sec. B of the SM).

We remark that, in both the $\rm{LL_{yz}}$ phase and   the ${\rm CF}_x$ phase, $|\varphi_k\rangle$ ($k=0,1,\cdots, L$) are simultaneous eigenvectors of the model Hamiltonian  (\ref{xyz2}) and $V_q$.
More precisely, we have $H|\varphi_k\rangle = E_k |\varphi_k\rangle$ and $V_q|\varphi_k\rangle = \exp(i2\pi k/q) |\varphi_k\rangle$ ($k=0,1,\cdots, L$). In this sense,   $|\varphi_k\rangle$ ($k=0,1,\cdots, L$)
constitute a one-dimensional representation for the cyclic group ${\rm Z}_{q}$.
In the thermodynamic limit $L \rightarrow \infty$,  ${\rm Z}_{q}$ becomes the symmetry group $\rm{U}(1)$ generated 
by $K_{yz}$. In addition, $|\varphi_k\rangle$ ($k=0,1,\cdots, L$) are always invariant under the symmetry group $\rm{U}(1)$ generated  
by $K_x$.

Apart from the commonalities for the $\rm{LL_{yz}}$ phase and   the ${\rm CF}_x$ phase, we are more interested in the differences between them. The emergence of  the cyclic group ${\rm Z}_{q}$  is one of the characteristic features in both the $\rm{LL_{yz}}$ phase and the ${\rm CF}_x$ phase. Actually,  the $L/2$-th and $L/2+1$-th low-lying states, which form the pair with the energy eigenvalue $J_x^2L$, take the form $|\otimes_l \{0_y)0_z \}_l\rangle$
and   $|\otimes_l \{0_z)0_y \}_l\rangle$, respectively, in the two phases.
However,  the gap $\Delta_{0L/2}$ survives in the $\rm{LL_{yz}}$ phase,  but vanishes in the ${\rm CF}_x$ phase, when the thermodynamic limit is approached. This observation makes it possible to distinguish the ${\rm CF}$ phases from the $\rm{LL}$ phases.

The essential difference manifests itself in the observation that
the ground state $|\varphi_0\rangle$ and $L$ low-lying states $|\varphi_k\rangle$ ($k=1,\cdots, L$) are quasi-degenerate in the ${\rm CF}_x$ phase, but not in the $\rm{LL_{yz}}$ phase.
The consequence will be elaborated on in the context of  H-orthogonal states~\cite{shiqq2} below. Physically, the H-orthogonal states offer a description for a finite-size precursor to the $\rm{U(1)}$ symmetry-broken states that {\it only} occur in the thermodynamic limit, thus leading to an alternative SSB pattern for $\rm{U(1)}$ in the $\rm{CF}_x$ phase. In contrast, this is not true for $|\varphi_k\rangle$ ($k=0,1,\cdots, L$) in the $\rm{LL_{yz}}$ phase, given that the low energy physics is described as a conformal field theory. 

Meanwhile, given a permutation symmetry group $S_{L/2}$ for the ground state subspace emerges on the  two characteristic lines $J_y/J_z=1$ and $J_x=0$ and they are  located in the $\rm{CF_{x}}$ phase,
one may anticipate that there exists an emergent permutation symmetry group $S_{L/2}$, away from the  two characteristic lines $J_y/J_z=1$ and $J_x=0$. As it turns out,  such an emergent permutation symmetry group $S_{L/2}$ for the ground state subspace is approximate for finite $L$'s, but becomes exact when $L$ tends to infinity (for more details, cf. Sec. D of the SM).

\section{H-orthogonality and  an alternative SSB pattern for $\rm{U(1)}$}

Given we are able to construct exactly the $q$ H-orthogonal states on the characteristic line $J_x=0$, it is plausible to
resort to a generic scheme on  H-orthogonal states~\cite{shiqq2} to characterize  an alternative SSB pattern for $\rm{U(1)}$ in the $\rm{CF}_x$ phase, away from the two characteristic lines $J_y/J_z=1$ and $J_x=0$.
As a result of  the cyclic group ${\rm Z}_{q}$, the Hilbert space is separated into disjoint sectors, labeled by the phases $w_k=\exp({i2\pi k/q})$, with $k=0, 1, 2,...,q-1$, 
as far as the low-energy physics is concerned.  More precisely, we construct $q$ H-orthogonal states $|\psi_k\rangle$ from the ground state $|\varphi_0\rangle$ and  the low-lying states $|\varphi_k\rangle$ ($k=1,\cdots, L$):
\begin{equation}
|\psi_k\rangle=\sum_{k'=0}^{q-1} c_{k'} (w_k)^{k'}|\varphi_{k'}\rangle.
\end{equation}
Here, $|\psi_k\rangle$ is normalized so that $\sum_k|c_k|^2=1$. However, they are not orthogonal to each other.
Instead,
$|\psi_k\rangle$ ($k=0, 1, 2,...,q-1$) satisfy the H-orthogonality~\cite{shiqq2}
\begin{eqnarray*}
\langle \psi_k'|H|\psi_{k}\rangle=0, {\rm if}\; k \neq k'.
\end{eqnarray*}
Then we have
\begin{equation}
 \begin{aligned}
|c_k|^2E_k=|c_{q-1}|^2E_{q-1},\\
|c_{q-1}|^2E_{q-1}=\frac{1}{\sum_{k=0}^{q-1}\frac{1}{E_k}}.
\end{aligned}
\end{equation}
From $E_l= E_{q-l}$ ($l=1,\cdots,L/2$), it follows that $|c_l|^2= |c_{q-l}|^2$. In fact, the low-lying state $|\varphi_l\rangle$ is mapped to the low-lying state $|\varphi_{q-l}\rangle$ and vice versa under the one-site translation operation if PBCs are adopted or  under the permutation $P_{12}P_{23}\cdots,P_{L-1,L}$ if OBCs are adopted. In particular, the ground state is invariant under the one-site translation operation if PBCs are adopted or under the permutation $P_{12}P_{23}\cdots,P_{L-1,L}$ if OBCs are adopted. 
We stress that the one-site translation operation is a symmetry operation under PBCs. However, it does not commute with the generators $K_{yz}$ and $K_{x}$ of the symmetry group ${\rm U}(1) \times {\rm U}(1)$.  Hence, it is impossible to simultaneously diagonalize the model 
Hamiltonian (\ref{xyz2}),  the generators $K_{yz}$ and $K_{x}$ of the symmetry group ${\rm U}(1) \times {\rm U}(1)$ and the one-site translation operation under PBCs. Here, we have chosen to simultaneously diagonalize the model 
Hamiltonian (\ref{xyz2}) and the generators $K_{yz}$ and $K_{x}$ of the symmetry group ${\rm U}(1) \times {\rm U}(1)$. 

A peculiar feature of the $q$ H-orthogonal states  $|\psi_k\rangle$  ($k=0, 1, 2,...,L$) is that they satisfy  a cyclic relation: $|\psi_{k+1}\rangle=V_q|\psi_k\rangle$.
This cyclic relation implies that the fidelity $\langle\psi_k'|\psi_k\rangle$ between any two H-orthogonal states $|\psi_k'\rangle$ and $|\psi_k\rangle$ scales with the system size $L$ exponentially, i.e.,
$\langle\psi_k'|\psi_k\rangle\equiv d_{kk'}^L$, with $d_{kk'}$ being the fidelity per lattice site~\cite{zhou}, which vanishes in the thermodynamic limit, if $k \neq k'$. Physically, the H-orthogonal states may be understood as a finite-size precursor to the symmetry-broken states arising from an alternative SSB pattern for ${\rm U}(1)$ that {\it only} occurs in the thermodynamic limit. However, it is necessary to require that the ground state $|\varphi_0\rangle$ and $L$ low-lying states $|\varphi_k\rangle$ ($k=1,\cdots, L$) are quasi-degenerate, in order to ensure the existence of the H-orthogonal states.

Conversely, the ground state $|\varphi_0\rangle$ and  the low-lying states $|\varphi_k\rangle$ ($k=1,\cdots, L$) may be expressed in terms of the $q$ H-orthogonal states $|\psi_k\rangle$
\begin{equation}
|\varphi_k\rangle=\frac{1}{c_k}\sum_{k'=0}^{q-1}(w_{k'})^{-k}|\psi_{k'}\rangle.
\label{inverse}
\end{equation}

Note that  the energy expectation value $\bar{E}_k$ for a H-orthogonal state, defined as $\langle\psi_k|\mathscr{H}|\psi_k\rangle$, is identical for any $k$: ${\bar  E(L)}_k= \bar{E(L)}$.
In fact, ${\bar E(L)}$ takes the form
\begin{equation}
{\bar E(L)}=\sum_{k=0}^{L} |c_k(L)|^2E_k(L) = E_0(L)+ \sum_{k=1}^L |c_k(L)|^2 \Delta_k(L),
\end{equation}
with $\Delta_k (L)= (E_k(L)-E_0(L))$ being the $k$-th energy gap. Note that $\sum_{k=1}^L |c_k(L)|^2 \Delta_k (L)= 2\sum_{l=1}^{L/2} |c_l(L)|^2 \Delta_l(L)$ .
If one expresses $|c_l(L)|^2$ in terms of $\langle\psi_{k'}|\psi_k\rangle$, then $\sum_{l=1}^{L/2} |c_l(L)|^2 \Delta_l(L)$ is split into two parts:
$\sum_{l=1}^{L/2} |c_l(L)|^2 \Delta_l(L) = 1/q \sum_{l=1}^{L/2} \Delta_l(L) + \sum_{l=1}^{L/2} (|c_l(L)|^2-1/q) \Delta_l(L)$. Here, $ 1/q \sum_{l=1}^{L/2} \Delta_l(L)
\approx 1/L \sum_{l=1}^{L/2} \Delta_l(L)$ is half the arithmetic average $\Delta_a(L)$ of the gaps  $\Delta_k (L)= (E_k(L)-E_0(L))$.
For fixed $L$,  $|c_l(L)|^2$ decreases and $\Delta_l(L)$ increases, as $l$ increases from 1 to $L/2$. Hence,  $|(c_l(L)|^2 -1/q)\Delta_l(L)$ is smoothly varying as $l$ increases.
Taking into account the fact that the fidelity between any two H-orthogonal states scales with the system size $L$ exponentially, and combining with an observation that both $|c_k(L)|^2 $ and $\Delta_k(L)$, as a function of $L$, are  monotonically decreasing with increasing $L$, we neglect $\sum_{l=1}^{L/2} (|c_l(L)|^2-1/q)\Delta_l(L)$, or equivalently, $\sum_{k=1}^L (|c_k(L)|^2 -1/q)\Delta_k (L)$, and
make a reasonable estimate 
$ \Delta_a(L) \approx B L \exp(-\kappa L)$, with $B$ and $\kappa$ being positive constants.
Hence, the ground state energy $E_0$ becomes
\begin{equation}
E_0(L)\approx {\bar E(L)}-B\;L e^{-\kappa L}.
\label{HO}
\end{equation}

We are led to conclude that, for a finite system size $L$, the ground state $|\varphi_0\rangle$ and  the low-lying states $|\varphi_k\rangle$ ($k=1,\cdots, L$) in  
the $\rm{CF}_x$ phase are quasi-degenerate, due to the presence of  $q$ H-orthogonal states $|\psi_k\rangle$, in sharp contrast to their counterparts in the $\rm{LL_{yz}}$  phase.
We remark that the $\rm{CF}_x$ phase becomes infinitely degenerate in the thermodynamic limit $L \rightarrow \infty$,  as a result of
an alternative SSB pattern for $\rm{U}(1)$, featuring that no gapless GM emerges. This is due to the fact that  this SSB pattern for a continuous symmetry group $\rm{U}(1)$ is {\it only} a limit of a SSS pattern for a discrete symmetry group ${\rm Z}_{q}$, when $q$ tends to  infinity. Indeed, infinitely degenerate ground states are (unentangled) factorized states so that they are scale-invariant. Actually, one may still introduce the fractal dimension $d_f$ to 
describe scale-invariant factorized states, with $d_f$ being zero~\cite{entropy}. In fact, 
the number of type-B GMs $N_B$, identical to the fractal dimension $d_f$,  must be zero, as follows from the counting rule of GMS~\cite{watanabe,watanabe2}, given only one generator $K_{yz}$ is broken.

As a result of this SSB pattern for $\rm{U}(1)$,  infinitely degenerate ground states, as (unentangled) factorized states, are permutation-invariant. Hence,
the emergent permutation symmetry $S_{L/2}$ (with respect to the unit cells consisting of the two nearest-neighbor lattice sites) in the ground state subspace, as already observed on the two characteristic lines $J_y/J_z=1$ and $J_x=0$, exist in the entire $\rm{CF_{x}}$ phase, when the thermodynamic limit is approached.

\section{Finite-size corrections to the ground state energy $E_0$: emergent permutation symmetry}
 
An asymptotic analysis, up to the first-order correction, is performed for the $q$ H-orthogonal states  $|\psi_k\rangle$  in Sec. E of the SM.  As a result,
the energy expectation value $\bar{E}_k$ for a H-orthogonal state takes the form 
\begin{equation}
{\bar E}(L) =J_x^2L-A e^{{\eta}/{L}},
\end{equation}
where $A=2a^2/\omega_0$ and $\eta= \omega_0 g/(2 a^2)$, with $g$ being a positive constant, and $a=J_x(J_z-J_y)$.
Hence, the ground state energy  $E_0(L)$ takes the form
\begin{equation}
E_0(L)={J_x}^2L -A e^{{\eta}/{L}} - BLe^{-\kappa L}.
\label{el}
\end{equation}
Physically,  two length scales are  competing with each other in the ${\rm CF}_x$ phase: one is involved in the second term originating from the emergent permutation symmetry in the ground state subspace and the other is involved in the third term, originating from an alternative SSB pattern for $\rm{U(1)}$. 

\begin{figure}
 	\includegraphics[width=0.3\textwidth]{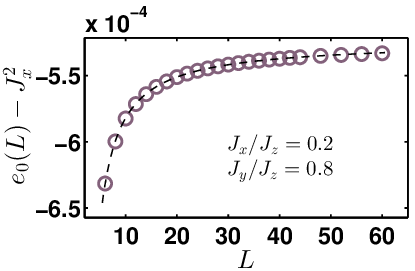}
 	\caption{(color online) The finite-size corrections to the ground state energy per lattice site, denoted as $e_0(L)$, for the point $(0.2,0.8)$  in the ${\rm CF}_x$ phase. Here, the finite-size DMRG algorithm is exploited to simulate the model (\ref{xyz2}) under PBCs, with the system size $L$ ranging from $6$ to $60$.
 	 The best fit yields that  $A=4.735\times10^{-4}$, $B=5.255\times 10^{-4}$, $\eta=1.8$, and $\kappa=0.2687\times 10^{-4}$ in Eq.(\ref{el}).
 }
 	\label{EEFigcf}
 \end{figure}

For a few randomly chosen points in the $\rm{CF_{x}}$ phase, it is found that the ground state energy $E_0(L)$, evaluated from the finite-size density matrix renormalization group (DMRG) simulations~\cite{DM1}, agrees well with the theoretical prediction in Eq.~\ref{el}, even for small $L$'s, with $L$ ranging from 6 to 60, as seen in Fig.~\ref{EEFigcf} from our simulation results for the point $(0.2,0.8)$ in the  $\rm{CF_{x}}$ phase (for more details, cf. Sec. F of the SM). This lends further support to our conclusion that the $\rm{CF_{x}}$ phase 
results from an alternative SSB pattern for the symmetry group $\rm{U(1)}$ generated from $K_{yz}$,  thus leading to
infinitely degenerate ground states that are scale-invariant in the thermodynamic limit, thus representing an exotic quantum state of matter.

The finite-size corrections to the ground state energy mark an essential difference between the $\rm{CF_{x}}$ phase and the $\rm{LL_{yz}}$ phase. The former is scale-invariant, but not conformally invariant, whereas the latter is conformally invariant, with central charge $c=1$, subject to  the finite-size corrections to the ground state energy predicted from conformal field theory~\cite{cardy}. 
 
\section{summary}

A systematic investigation has been carried out for an anisotropic extension of the ferromagnetic spin-1 biquadratic model, in an attempt to characterize the three distinct types of phases -  the $\rm{SPt}$ phases, the $\rm{LL}$ phases, and  the $\rm{CF}$ phases. As it turns out, the $\rm{SPt}$ phases are gapful, whereas the  $\rm{LL}$ phases  are gapless and conformally invariant.  In contrast, the $\rm{CF}$ phases represent an exotic quantum state of matter, which results from an alternative SSB pattern for $\rm{U}(1)$, with a salient feature that no gapless GM emerges. Meanwhile,
such an alternative SSB pattern for $\rm{U}(1)$ yields infinitely degenerate ground states in the thermodynamic limit, which turn out to be (unentangled) factorized states. Hence, they are scale-invariant, with the fractal dimension $d_f$, identical to  the number of type-B GMs $N_B$, being zero~\cite{entropy}.

In conclusion, the presence of an alternative SSB pattern for $\rm{U(1)}$  does not challenge the counting rule of GMs for continuous SSB, since no GM emerges. However, it does require clarification of the semantic meaning for continuous SSB in the Goldstone theorem~\cite{GM}, since $\rm{U}(1)$, as a continuous symmetry group,  makes an exceptional case, simply due to the well-known mathematical result that a continuous symmetry group $\rm{U}(1)$ may be regarded as a limit of a discrete symmetry group $Z_q$, when $q$ tends to infinity. That is, the dichotomy between continuity and discreteness is more involved than one might have expected.

\section{Acknowledgements}

We are grateful to Murray Batchelor, John Fjaerestad and Ian McCulloch for comments and suggestions to improve the manuscript.

\newpage
\onecolumngrid
\newpage
\section*{Supplementary Material}
\twocolumngrid
\setcounter{page}{1}
\setcounter{equation}{0}
\setcounter{figure}{0}
\renewcommand{\theequation}{S\arabic{equation}}
\renewcommand{\thefigure}{S\arabic{figure}}
\renewcommand{\bibnumfmt}[1]{[S#1]}
\renewcommand{\citenumfont}[1]{S#1}

\subsection{Some exact results on the two characteristic lines:	$J_y/J_z=1$ and $J_x/J_z=0$}\label{twolines}

For our purpose, we present some exact results in details on the two characteristic lines: 
$J_y/J_z=1$ with $J_x/J_z<1$ and $J_x/J_z=0$ with $J_y/J_z > 0$.

On the characteristic line: $J_y/J_z=1$ with $J_x/J_z<1$, there is one symmetry group ${\rm SU(2)}$, generated by
$\Sigma_x=\sum_j(-1)^{j+1}(S_j^yS_j^z+S_j^zS_j^y)/2$, $\Sigma_y=\sum_jS_j^x/2$, and $\Sigma_z=K_{yz}/2$, together with one symmetry group $\rm{U(1)}$, generated by $K_{x}=\sum _j (-1)^{j+1}(S_j^x)^2$. A SSB pattern from $\rm{SU(2)}\times \rm{U(1)}$ to  $\rm{U(1)}\times \rm{U(1)}$ arises, thus leading to
highly degenerate ground states, with the ground state energy  $E_0$ being equal to $J_x^2L$.
Two of the degenerate ground states are invariant under the one-site translation operation:
$|\otimes_{j=1}^L\{1_x\}_j\rangle$ and $|\otimes_{j=1}^L\{-1_x\}_j\rangle$, where $|1_x\rangle$ and $|-1_x\rangle$ are the eigenvectors of $S_j^x$, with the eigenvalues being $1$ and $-1$, respectively.
Actually, $|\otimes_{j=1}^L\{1_x\}_j\rangle$ and $|\otimes_{j=1}^L\{-1_x\}_j\rangle$ are the highest and lowest weight states for the symmetry group  ${\rm SU(2)}$ in the $\Sigma_y$ representation. Indeed, a sequence of degenerate ground states $|L,M\rangle_y$ are generated from the repeated action of the lowering operator $\Sigma_-$ of the symmetry group  ${\rm SU(2)}$ in the $\Sigma_y$ representation on the highest weight state $|\otimes_{j=1}^L\{1_x\}_j\rangle$:  $|L,M\rangle_y=1/\sqrt{C_L^M}\Xi_-^M |\otimes_{j=1}^L\{1_x\}_j\rangle$.
Meanwhile, $|\otimes_{j=1}^{L/2}\{0_z0_y\}_j\rangle$ and  $|\otimes_{j=1}^{L/2}\{0_y0_z\}_j\rangle$ are the highest and lowest weight states for the symmetry group  ${\rm SU(2)}$
in the $\Sigma_z$ representation.
Therefore, a sequence of degenerate ground state $|L,M\rangle_z$ generated from the repeated action of the lowering operator $\Sigma_-$ of the symmetry group  ${\rm SU(2)}$ in the $\Sigma_z$ representation on  the highest weight state $|\otimes_{j=1}^{L/2}\{0_z0_y\}_j\rangle$:
$|L,M\rangle_z\equiv \frac{1}{\sqrt{C_L^M}} \Sigma_-^M |\otimes_{j=1}^{L/2}\{0_z0_y\}_j\rangle$. 
As shown in Refs.~\cite{smFMGM,smgolden}, the entanglement entropy $S(L,N)$ for this type of degenerate ground states arising from SSB with type-B GMs exhibits a logarithmic scaling relation with the block size $N$ in the thermodynamic limit $L \rightarrow \infty$, with the prefactor being half the number of type-B GMs $N_B$: $N_B=1$, as long as the filling $f\equiv M/L$ is non-zero.

On the characteristic line  $J_x/J_z=0$ with $J_y/J_z > 0$, there exists a two-parameter family of factorized ground states $|\phi_f(L)\rangle=\bigotimes_l |v_1v_2\rangle _l$~\cite{smspin1BM}, where $|v_1v_2\rangle _l = |v_1\rangle_{2l-1} |v_2\rangle _{2l}$,
with $|v_1 \rangle_{2l-1}$ and  $|v_2 \rangle_{2l}$ being given in  Eq.~(\ref{v12}). 
The number of linearly independent ground states is $q=L+1$. We remark that there are {\it only} two factorized ground states $|\phi_0\rangle$ and  $|\phi_{a0}\rangle$, invariant under the one-site translation operation: one is
$|\phi_0\rangle=\bigotimes_j |v\rangle_j$, where $|v \rangle_j=\mu |0_y\rangle_j + \nu |0_z\rangle_j$, with $\mu=\sqrt {J_y/(J_y+J_z)}$ and $\nu=\sqrt {J_z/(J_y+J_z)}$, and the other
is  $|\phi_{a0}\rangle=\bigotimes_j |v\rangle_j$, where $|v \rangle_j=\mu |0_y\rangle_j + \nu |0_z\rangle_j$, with $\mu=\sqrt {J_y/(J_y+J_z)}$ and $\nu=-\sqrt {J_z/(J_y+J_z)}$.
Note that  $|\phi_0\rangle$ and  $|\phi_{a0}\rangle$ evolve into $|\otimes_{j=1}^L\{1_x\}_j\rangle$ and $|\otimes_{j=1}^L\{-1_x\}_j\rangle$, as $J_y$ tends to $J_z$.
From both  $|\phi_{0}\rangle$ and  $|\phi_{0a}\rangle$, we may construct two sequences of the $q$ H-orthogonal states:  $|\psi_k\rangle \equiv V_q^k |\phi_{0}\rangle$ and $|\psi_{ak}\rangle \equiv V_q^k |\phi_{a0}\rangle$. According to Eq.(\ref{inverse}), we may express the ground state $|\varphi_0\rangle$ and $L$ low-lying states $|\varphi_k\rangle$, with $k=1,\cdots, L$ 
in terms of the $q$ H-orthogonal states $|\psi_k\rangle$ or $|\psi_{ak}\rangle$. In particular,  the ground state $|\varphi_0\rangle$ takes the form: $|\varphi_0\rangle=\frac{1}{c_0}\sum_{k'=0}^{q-1}|\psi_{k'}\rangle$, which in turn allows us to expand it into a linear combination of the highly degenerate ground states. Therefore, the 
entanglement entropy $S(L,N)$ for  the ground state $|\varphi_0\rangle$ may be evaluated.  In fact, the ground state $|\varphi_0\rangle$ on the characteristic line  $J_x/J_z=0$ with $J_y/J_z > 0$ is highly entangled.

\subsection{The ground state and the low-lying states from an exact diagonalization perspective}\label{Lowlyingstate}

For the spin-1 ferromagnetic anisotropic biquadratic model (\ref{xyz2}) under PBCs, the ED simulations are performed
for six chosen points $(J_x/J_z, J_y/J_z)=(0.35, 0.94), (0.3, 0.94), (0.2, 0.9), (0.2, 0.8), (0.15, 0.8)$ and $(0.1, 0.7)$ in the ${\rm CF}_x$ phase, 
and four chosen points $(J_x/J_z, J_y/J_z)=(0.25, 0.65),  (0.3, 0.7), (0.35, 0.7)$ and  $(0.4, 0.75)$ in the  $\rm{LL_{yz}}$ phase, respectively.

We target at the ground state $|\varphi_0\rangle$ and the $L$ low-lying states $|\varphi_{k}\rangle$ ($k=1,\cdots,L$), with the ground  state energy $E_0$ and the 
energy eigenvalues $E_k$ for each of the chosen points in the two phases.
We take the point $(J_x/J_z=0.2, J_y/J_z=0.9)$ in the ${\rm CF}_x$ phase and the point $(J_x/J_z=0.3, J_y/J_z=0.7)$ in the $\rm{LL_{yz}}$ phase, with the system size $L=4, 8$ and $16$, as two illustrative examples.
The ground state energy per lattice site, denote as $e_0 \equiv E_0/L$, and the energy eigenvalues per lattice site, denote as $e_k \equiv E_k/L$,
are listed in Table~\ref{table1} and  in Table~\ref{tableLL}, respectively. The numerical results clearly show that the $L$ low-lying states $|\varphi_{k}\rangle$  occur in pair, in addition to the non-degenerate 
ground state $|\varphi_{0}\rangle$ for even $L$'s. The same pattern shows up at
the other five points in the ${\rm CF}_x$  phase and at the three other points in the $\rm{LL_{yz}}$ phase  (but are not shown here).

Since the symmetry group ${\rm U}(1) \times {\rm U}(1)$ is not implemented during our numerical simulations,  the $L$ low-lying states $|\varphi_k\rangle$ ($k=1,\cdots, L$)
are not eigenvectors of the  ${\rm U}(1)$ symmetry generator $K_{yz}$. Hence, it is necessary to take one additional step to ensure the symmetry group ${\rm U}(1)$ generated 
from $K_{yz}$. For our purpose, we take advantage of the operator
$V_q=\prod _j\exp(i2\pi K_{yz}^j/q)$, with $q=L+1$, to introduce  $2 \times 2$ matrix $M_{l,q-l}$
for each pair of $|\varphi_{l}\rangle$ and $|\varphi_{q-l}\rangle$ ($l=1,\cdots,L/2$):
\begin{equation}
	M_{l,q-l}=
	\left (
	\begin{matrix}
		\langle \varphi_{l}|V|\varphi_{l}\rangle&\langle \varphi_{l}|V|\varphi_{q-l}\rangle\\
		\langle \varphi_{q-l}|V|\varphi_{l}\rangle&\langle \varphi_{q-l}|V|\varphi_{q-l}\rangle
	\end{matrix}
	\right).
\end{equation}
Then, the diagonalization of the matrix $M_{l,q-l}$  yields simultaneous eigenvectors of the model Hamiltonian  (\ref{xyz2}) and $V_q$, which are listed in Table~\ref{table2}..
Here, we remark that $\langle \varphi_{l}|V|\varphi_{q-l}\rangle$,  defined as the elements of $V_q$ between the two degenerate low-lying states $|\varphi_{l}\rangle$ and $|\varphi_{q-l}\rangle$, are evaluated. From now on, we use the same notations to denote the $L$ low-lying states $|\varphi_{m}\rangle$ that are simultaneous eigenvectors of the model Hamiltonian  (\ref{xyz2}) and $V_q$.

Hence, we conclude that the $l$-th low-lying states $|\varphi_{l}\rangle$ is degenerate with $q-l$-th low-lying states $|\varphi_{q-l}\rangle$, in both  the ${\rm CF}_x$ phase and  the $\rm{LL_{yz}}$ phase, i.e., $E_l=E_{q-l}$ ($l=1,\dots, L/2)$. Meanwhile, $|\varphi_{l}\rangle$ is mapped to $|\varphi_{q-l}\rangle$ and vice versa under the one-site translation operation, given PBCs are adopted.

We remark that the same pattern is not valid in the $\rm{SPt_{z}}$ phase. The ground state energy per lattice site $e_0(L)$ and the energy eigenvalues per lattice site, denoted as $e_k(L)$ ($k=1,\cdots,L$) are listed in Table~\ref{tableSPt},  for the $L$ low-lying states, at the point  $(J_x/J_z=0.4, J_y/J_z=0.47)$ in the $\rm{SPt_{z}}$ phase.
This makes it possible to distinguish the $\rm{SPt_{z}}$ phase from both  the ${\rm CF}_x$ phase and the $\rm{LL_{yz}}$ phase.

\begin{table*}
	\renewcommand\arraystretch{1.5}
	\caption{The ground state energy per lattice site $e_0(L)$ and the energy eigenvalues per lattice site, denoted as $e_k(L)$ ($k=1,\cdots,L$), for the $L$ low-lying states, respectively, with the system size $L=4, 8$ and $16$, at the point $(J_x/J_z=0.2, J_y/J_z=0.9)$ in the ${\rm CF}_x$ phase.}
	\begin{tabular}{c|ccccccccccccccccc}
		\hline
		\hline
		&$e_0(L)$ & $e_1(L)$ & $e_2(L)$ & $e_3(L)$ & $e_4(L)$ & $e_5(L)$ & $e_6(L)$ & $e_7(L)$ & $e_8(L)$ & \\
		\hline
		\begin{minipage}{1.1cm} $L=4$ \end{minipage}
		& 0.03984448& 0.03988474 & 0.03988474&0.04 & 0.04&\\
		\begin{minipage}{1.1cm} $L=8$ \end{minipage}
		&0.03986720& 0.03987561 & 0.03987561 &0.03990076 & 0.03990076& 0.03994237 &0.03994237& 0.04& 0.04&\\
		\hline
		&$e_0(L)$ & $e_1(L)$ & $e_2(L)$ & $e_3(L)$ & $e_4(L)$ & $e_5(L)$ & $e_6(L)$ & $e_7(L)$ & $e_8(L)$ & \\
		\begin{minipage}{1.1cm} $L=16$ \end{minipage}
		&0.03987611 &0.03987807 & 0.03987807& 0.03988393 & 0.03988393 &0.03989368& 0.03989368 &0.03990732& 0.03990732& \\
		&$e_9(L)$ & $e_{10}(L)$ & $e_{11}(L)$ & $e_{12}(L)$ & $e_{13}(L)$ & $e_{14}(L)$ & $e_{15}(L)$ & $e_{16}(L)$& \\
		&0.03992480& 0.03992480 &0.03994610 &0.03994610&0.03997119 &0.03997119& 0.04 &0.04 &\\
		\hline
		\hline
	\end{tabular}
	\label{table1}
\end{table*}

\begin{table*}
	\renewcommand\arraystretch{1.5}
	\caption{The ground state energy per lattice site $e_0(L)$ and the energy eigenvalues per lattice site, denoted as $e_k(L)$ ($k=1,\cdots,L$), for the $L$ low-lying states, respectively, with the system size $L=4, 8$ and $16$, at the point $(J_x/J_z=0.3, J_y/J_z=0.7)$ in the ${\rm LL}_{yz}$ phase.}
	\begin{tabular}{c|ccccccccccccccccc}
		\hline
		\hline
		&$e_0(L)$ & $e_1(L)$ & $e_2(L)$ & $e_3(L)$ & $e_4(L)$ & $e_5(L)$ & $e_6(L)$ & $e_7(L)$ & $e_8(L)$ & \\
		\hline
		\begin{minipage}{1.1cm} $L=4$ \end{minipage}
		& 0.08541595& 0.08702758 & 0.08702758&0.09 & 0.09&\\
		\begin{minipage}{1.1cm} $L=8$ \end{minipage}
		&0.08622324&0.08650791 & 0.08650791&0.08730797& 0.08730797&0.08851379 &0.08851379&0.09&0.09&\\
		\hline
		&$e_0(L)$ & $e_1(L)$ & $e_2(L)$ & $e_3(L)$ & $e_4(L)$ & $e_5(L)$ & $e_6(L)$ & $e_7(L)$ & $e_8(L)$ & \\
		\begin{minipage}{1.1cm} $L=16$ \end{minipage}
		&0.08648798& 0.08655110&  0.08655110& 0.08673843& 0.08673843& 0.08704421& 0.08704421&  0.08745986&0.08745986& \\
		&$e_9(L)$ & $e_{10}(L)$ & $e_{11}(L)$ & $e_{12}(L)$ & $e_{13}(L)$ & $e_{14}(L)$ & $e_{15}(L)$ & $e_{16}(L)$& \\
		&0.08797491& 0.08797491& 0.08857788& 0.08857788& 0.08925689& 0.08925689& 0.09& 0.09& \\
		\hline
		\hline
	\end{tabular}
	\label{tableLL}
\end{table*}

\begin{table*}
	\renewcommand\arraystretch{1.2}
	\caption{The eigenvalues of the matrix $M_{l,q-l}$, defined as the elements of $V_q$ between the two degenerate low-lying states $|\varphi_{l}\rangle$ and $|\varphi_{q-l}\rangle$, with the system size $L=4, 8$ and $16$, at the point $(J_x/J_z=0.2, J_y/J_z=0.9)$ in the ${\rm CF}_x$ phase and at the point $(J_x/J_z=0.3, J_y/J_z=0.7)$ in the ${\rm LL}_{yz}$ phase.}
	\begin{tabular}{c|c|c|c|c|c|c|c|cccccccccc}
		\hline
		\hline
		& \begin{minipage}{1.8cm} $l$, $q-l$  \end{minipage} &
		\begin{minipage}{1.8cm}  $l$, $q-l$ \end{minipage} &
		\begin{minipage}{1.8cm}  $l$, $q-l$  \end{minipage} &
		\begin{minipage}{1.8cm}  $l$, $q-l$  \end{minipage} &
		\begin{minipage}{1.8cm} $l$, $q-l$  \end{minipage} &
		\begin{minipage}{1.8cm}  $l$, $q-l$  \end{minipage} &
		\begin{minipage}{1.8cm}  $l$, $q-l$   \end{minipage}&
		\begin{minipage}{1.8cm}  $l$, $q-l$  \end{minipage} \\
		& 1,$~~~$ $L$ & 2,$~~~$ $L-1$ & 3,$~~~$ $L-2$ & 4, $~~~$ $L-3$ & 5,$~~~$ $L-4$ & 6,$~~~$ $L-5$ & 7, $~~~$ $L-6$ & 8, $~~~$ $L-7$ &\\
		\hline
		\begin{minipage}{1cm} $L=4$ \end{minipage}
		&$e^{i2\pi \times 1/(L+1)}$ &$e^{i2\pi \times 2/(L+1)}$  & &&&&&&&\\
		&$e^{i2\pi \times 4/(L+1)}$ & $e^{i2\pi \times 3/(L+1)}$ & &&&&&&&\\
		\hline
		\begin{minipage}{1cm} $L=8$ \end{minipage}
		&$e^{i2\pi \times 1/(L+1)}$ &$e^{i2\pi \times 2/(L+1)}$  &$e^{i2\pi \times 3/(L+1)}$ &$e^{i2\pi \times 4/(L+1)}$&&&&&\\
		&$e^{i2\pi \times 8/(L+1)}$ & $e^{i2\pi \times 7/(L+1)}$ &$e^{i2\pi \times 6/(L+1)}$ &$e^{i2\pi \times 5/(L+1)}$&&&&&\\
		\hline
		\begin{minipage}{1cm} $L=16$ \end{minipage}
		&$e^{i2\pi \times 1/(L+1)}$ &$e^{i2\pi \times 2/(L+1)}$  & $e^{i2\pi \times 3/(L+1)}$&$e^{i2\pi \times 4/(L+1)}$&$e^{i2\pi \times 5/(L+1)}$&$e^{i2\pi \times 6/(L+1)}$&$e^{i2\pi \times 7/(L+1)}$&$e^{i2\pi \times 8/(L+1)}$&\\
		&$e^{i2\pi \times 16/(L+1)}$ & $e^{i2\pi \times 15/(L+1)}$ & $e^{i2\pi \times 14/(L+1)}$&$e^{i2\pi \times 13/(L+1)}$&$e^{i2\pi \times 12/(L+1)}$&$e^{i2\pi \times 11/(L+1)}$&$e^{i2\pi \times 10/(L+1)}$&$e^{i2\pi \times 9/(L+1)}$&&\\
		\hline
		\hline
	\end{tabular}
	\label{table2}
\end{table*}

\begin{table*}
	\renewcommand\arraystretch{1.5}
	\caption{The ground state energy per lattice site $e_0(L)$ and the energy eigenvalues per lattice site, denoted as $e_k(L)$ ($k=1,\cdots,L$), for the $L$ low-lying states, respectively, with the system size  $L=4, 8, 12$ and $16$, at the point $(J_x/J_z=0.4, J_y/J_z=0.47)$ in the ${\rm SPt}_{z}$ phase.}
	\begin{tabular}{c|ccccccccccccccccc}
		\hline
		\hline
		&$e_0(L)$ & $e_1(L)$ & $e_2(L)$ & $e_3(L)$ & $e_4(L)$ & $e_5(L)$ & $e_6(L)$ & $e_7(L)$ & $e_8(L)$ & \\
		\hline
		\begin{minipage}{1.1cm} $L=4$ \end{minipage}
		&0.11137607& 0.13961933&  0.13961933& 0.16& 0.16& \\
		\begin{minipage}{1.1cm} $L=8$ \end{minipage}
		&0.11714763& 0.12690821&  0.12690821& 0.13834780& 0.13834780& 0.14795967& 0.14795967&  0.14795967& 0.14795967&\\
		\hline
		&$e_0(L)$ & $e_1(L)$ & $e_2(L)$ & $e_3(L)$ & $e_4(L)$ & $e_5(L)$ & $e_6(L)$ & $e_7(L)$ & $e_8(L)$ & \\
		\begin{minipage}{1.1cm} $L=12$ \end{minipage}
		&0.11775204& 0.12379340&  0.12379340& 0.13087058& 0.13087058& 0.13184427& 0.13184427&  0.13184427& 0.131844273&\\
		&$e_9(L)$ & $e_{10}(L)$ & $e_{11}(L)$ & $e_{12}(L)$ & \\
		&0.13639967& 0.13799444& 0.13799444& 0.13799444&\\
		\hline
		&$e_0(L)$ & $e_1(L)$ & $e_2(L)$ & $e_3(L)$ & $e_4(L)$ & $e_5(L)$ & $e_6(L)$ & $e_7(L)$ & $e_8(L)$ & \\
		\begin{minipage}{1.1cm} $L=16$ \end{minipage}
		&0.11786452& 0.12230268&  0.12230268& 0.12622999& 0.12622999& 0.12622999& 0.12622999&  0.12737090& 0.12737090& \\
		&$e_9(L)$ & $e_{10}(L)$ & $e_{11}(L)$ & $e_{12}(L)$ & $e_{13}(L)$ & $e_{14}(L)$ & $e_{15}(L)$ & $e_{16}(L)$& \\
		&0.12962196& 0.13069791& 0.13069791& 0.13100286& 0.13100286& 0.13100286& 0.13100286& 0.13271958& \\
		\hline
		\hline
	\end{tabular}
	\label{tableSPt}
\end{table*}

In the $\rm{CF_{x}}$ phase and the $\rm{LL_{yz}}$ phase, the energy gaps $\Delta_{0L/2}(L)$, defined as $\Delta_{0L/2}(L)=E_{L/2}(L)-E_0(L)$, are scrutinized from an ED perspective, 
where $E_0(L)$ is the ground state energy per lattice site, and $E_{L/2}(L)$ is the energy eigenvalue per lattice site for the $L/2$-th low-lying state.

\begin{figure}
	\includegraphics[width=0.48\textwidth]{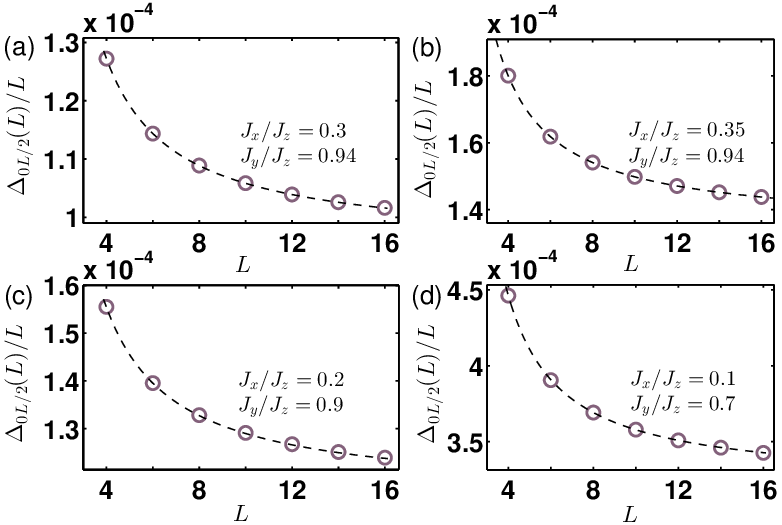}
	\caption{(color online) The ratio $\Delta_{0L/2}(L)/L$ as a function of the system size $L$ for four chosen points in the ${\rm CF}_x$ phase, with $L$ ranging from $4$ to $16$.
	}
	\label{CFFig1}
\end{figure}

In Fig.~\ref{CFFig1}, we plot the ratio $\Delta_{0L/2}(L)/L$ as a function of the system size $L$ for four chosen points: (a) $(J_x/J_z, J_y/J_z)=(0.3, 0.94)$, (b) $(J_x/J_z, J_y/J_z)=(0.35, 0.94)$, (c) $(J_x/J_z, J_y/J_z)=(0.2, 0.9)$, and
(d) $(J_x/J_z, J_y/J_z)=(0.1, 0.7)$
in the $\rm{CF_{x}}$ phase. The ED simulations are performed for a few different values of the system size $L$, ranging from $4$ to $16$.
According to Eq.~(\ref{el}), the energy gap $\Delta_{0L/2}(L)$ scales with the system size $L$ as
$\Delta_{0L/2}(L)=A\exp(\eta/L)+BL \exp(-\kappa L)$. Our simulation results for $A$ and $B$, $\eta$ and $\kappa$ are listed in Table~\ref{tablegapcf}. This amounts to performing an analysis of the finite-size corrections to the ground state energy $E_0$ for small $L$'s, accessible to the ED simulations. 
Although the results slightly deviate from those for larger $L$'s, they still indicate that $\Delta_{0L/2}(L)/L$ vanishes, as $L$ gets large.

\begin{figure}
	\includegraphics[width=0.48\textwidth]{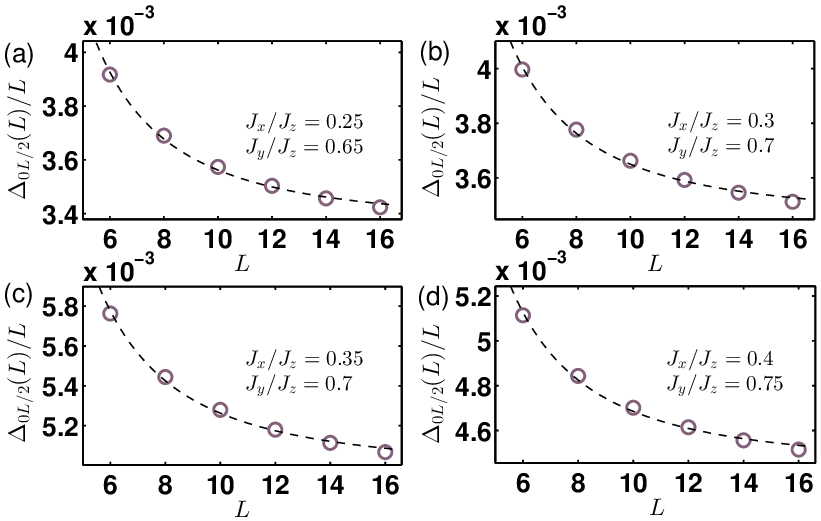}
	\caption{(color online) The energy gap $\Delta_{0L/2}(L)$ as a function of the system size $L$ for four chosen points in the ${\rm LL}_{yz}$ phase, with $L$ ranging from $4$ to $16$.
	}
	\label{LLFig1}
\end{figure}

In Fig.~\ref{LLFig1}, we plot the ratio $\Delta_{0L/2}(L)/L$ as a function of the system size $L$ for four chosen points: (a) $(J_x/J_z, J_y/J_z)=(0.25, 0.65)$, (b) $(J_x/J_z, J_y/J_z)=(0.3, 0.7)$, (c) $(J_x/J_z, J_y/J_z)=(0.35, 0.7)$, and
(d) $(J_x/J_z, J_y/J_z)=(0.4, 0.75)$
in the $\rm{LL_{yz}}$ phase. The energy gap $\Delta_{0L/2}(L)$ scales with the system size $L$ as
$\Delta_{0L/2}(L) \sim L$. Our simulation results indicate that $\Delta_{0L/2}(L)/L$ does not vanish, as $L$ goes to $\infty$.

\begin{table}
	\renewcommand\arraystretch{1.5}
	\caption{The parameters $A$, $B$, $\eta$ and $\kappa$ are extracted from the energy gap $\Delta_{0L/2}(L)=A \exp(\eta/L)+B\exp(-\kappa L)L$ for  the  model (\ref{xyz2}) under PBCs, with the system size $L$ ranging from $4$ to $16$.    }
	\begin{tabular}{c|ccccccc}
		\hline
		\hline
		&$J_x/J_z=0.3$&$J_x/J_z=0.35$&$J_x/J_z=0.2$&$J_x/J_z=0.1$&\\
		&$J_y/J_z=0.94$&$J_y/J_z=0.94$&$J_y/J_z=0.9$&$J_y/J_z=0.7$&\\
		\hline
		\begin{minipage}{1cm} $A$ \end{minipage}
		& $0.888\times 10^{-4}$ & $1.212\times 10^{-4}$ & $1.104\times 10^{-4}$ &$2.767\times 10^{-4}$&\\
		\begin{minipage}{1cm} $B$ \end{minipage}
		& $0.958\times 10^{-4}$ & $1.362\times 10^{-4}$ & $1.163\times 10^{-4}$ &$3.234\times 10^{-4}$&\\
		\begin{minipage}{1cm} $\eta$ \end{minipage}
		&$1.4$ & $1.5$ & $1.4$ &$2.3$&\\
		\begin{minipage}{1cm} $\kappa$ \end{minipage}
		& $0.821\times 10^{-4}$ &$3.169\times 10^{-4}$ & $0.609\times 10^{-4}$ & $0.999\times 10^{-4}$ &\\
		\hline
		\hline
	\end{tabular}
	\label{tablegapcf}
\end{table}

\begin{table}
	\renewcommand\arraystretch{1.5}
	\caption{The parameters $C$ and $D$ are extracted from the ratio $\Delta_{0L/2}(L)/L$, for  the  model (\ref{xyz2}), with the system size $L$ ranging from $6$ to $16$. Here, we assume that $\Delta_{0L/2}(L)/L= C + D/L^2$.}
	\begin{tabular}{c|ccccccc}
		\hline
		\hline
		&$J_x/J_z=0.25$&$J_x/J_z=0.3$&$J_x/J_z=0.35$&$J_x/J_z=0.4$&\\
		&$J_y/J_z=0.65$&$J_y/J_z=0.7$&$J_y/J_z=0.7$&$J_y/J_z=0.75$&\\
		\hline
		\begin{minipage}{1.3cm} $C$ \end{minipage}
		& $3.357\times 10^{-3}$ & $3.448\times 10^{-3}$ & $4.974\times 10^{-3}$ &$4.437\times 10^{-3}$&\\
		\begin{minipage}{1.3cm} $D$ \end{minipage}
		& $2.051\times 10^{-2}$ & $2.012\times 10^{-2}$ & $2.888\times 10^{-2}$ &$2.485\times 10^{-2}$&\\
		\hline
		\hline
	\end{tabular}
	\label{tablegapLL}
\end{table}

We remark that the $L/2-$th and $L/2+1-$th low-lying states are degenerate, thus forming a pair, with the energy eigenvalue per lattice site, denoted as $e_{L/2}$, being equal to the maximum $J_x^2$ 
among the $L$ low-lying states $|\varphi_{k}\rangle$ ($k=1,\dots,L$) in both the
$\rm{LL_{yz}}$ phase and the $\rm{CF_{x}}$ phase. Actually, the two degenerate low-lying states are identified as $|\otimes_l \{0_y)0_z \}_l\rangle$ and $|\otimes_l \{0_z)0_y \}_l\rangle$ in the two phases. This observation offers compelling evidence for our conclusion that
the ground state $|\varphi_0\rangle$ and $L$ low-lying states $|\varphi_k\rangle$ ($k=1,\cdots, L$) are quasi-degenerate in the ${\rm CF}_x$ phase, but not in the $\rm{LL_{yz}}$ phase.
In fact,  the symmetry-broken ground states arise from an alternative SSB pattern for ${\rm U}(1)$ that {\it only} occurs in the thermodynamic limit, with the ground state energy per lattice site being $J_x^2$ in the
${\rm CF}_x$ phase. Hence, they are (unentangled) factorized states such that the H-orthogonal states appear as a finite-size precursor to the symmetry-broken ground states. In contrast, this does not happen in the  $\rm{LL_{yz}}$ phase, since the ground state energy per lattice site is not equal to $J_x^2$. In  other words, the ratio $\Delta_{0L/2}(L)/L$ vanishes
in the ${\rm CF}_x$ phase, but it does not vanish in the  $\rm{LL_{yz}}$ phase.

\subsection{The entanglement entropy $S(L,N)$ in the $\rm{SPt_{z}}$, $\rm{LL_{yz}}$ and $\rm{CF_{x}}$ phases}\label{energygap}

The entanglement entropy $S(L,N)$ is investigated to characterize the three distinct phases - the $\rm{SPt_{z}}$, $\rm{LL_{yz}}$ and $\rm{CF_{x}}$ phases.

\begin{figure}[htbp]
	\includegraphics[width=0.47\textwidth]{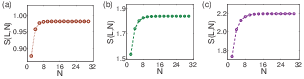}
	\caption{(color online) The entanglement entropy $S(L,N)$, as a function of the block size $N$, for the three points $(J_x/J_z, J_y/J_z)=$ (a) $(0.2, 0.3)$, (b) $(0.6, 0.65)$ and (c) $(0.8, 0.82)$  in the $\rm{SPt_{z}}$ phase. Here, the system size $L$ is 60. }
	\label{SPtFig1}
\end{figure}

In Fig.~\ref{SPtFig1}, we plot the  entanglement entropy $S(L,N)$, as a function of the block size $N$, for the three points $(J_x/J_z, J_y/J_z)=(0.2, 0.3), (0.6, 0.65)$ and $(0.8, 0.82)$  in the $\rm{SPt_{z}}$ phase. 
It is found that the entanglement entropy $S(L,N)$  saturates, as the block size $n$ increases. Here, the system size $L$ is chosen to be $L=60$. This indicates that an energy gap opens in this phase.

\begin{figure}[htbp]
	\vspace{4mm}
	\includegraphics[width=0.49\textwidth]{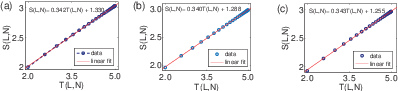}
	\caption{ The entanglement entropy $S(L,N)$ versus the block size $T(L,N)$ for three chosen points: (a) $(J_x/J_z,J_y/J_z) = (0.45, 0.75)$; (b) $(J_x/J_z,J_y/J_z) = (0.6, 0.75)$;  (c)  $(J_x/J_z,J_y/J_z) =(0.75, 0.85)$ in the ${\rm LL_{yz}}$ phase. Here,  $L=100$ and the bond dimension $\chi=40$. 
	}\label{Sfinite}
\end{figure}

As for the ${\rm LL}_{yz}$ phase, we demonstrate that it is gapless and its low energy physics is described in terms of conformal field theory. To this end,  we extract central charge $c$ from a finite-size scaling analysis of  the entanglement entropy $S(L,N)$. According to the conformal field theory~\cite{smcft}, $S(L,N)$ scales as
\begin{equation}
	S(L,N)=\frac{c}{3}T(L,N)+S_0,
\end{equation}
with $T(L,N)=\log_2{[L/\pi\sin(\pi N/L)]}$, where $c$ is central charge, and $S_0$ is a (model-dependent) additive constant.
For three chosen points  $(J_x/J_z, J_y/J_z)= (0.45, 0.75)$,  $(0.6, 0.75)$ and $(0.75, 0.85)$ in the ${\rm LL}_{yz}$ phase, we take advantage of  the variational finite-size MPS algorithm~\cite{smfrank} to simulate the model~(\ref{xyz2}) under PBCs.
In Fig.~\ref{Sfinite}, the entanglement entropy $S(L,N)$ versus $T(L,N)$, with the system size $L=100$ and the bond dimension $\chi=40$. 
In Table~\ref{tab2}, we list our simulation results for central charge $c$, which are  close to the exact value $c=1$, with the relative errors less than $ 3\%$.

\begin{table}[htbp]
	\centering
	\caption{Central charge $c$ extracted from the entanglement entropy $S(L,N)$ versus $T(L,N)$ for three chosen points in the $\rm{LL_{yz}}$ phase.  Our simulation results are  close to the exact value $c=1$, with the relative errors less than $ 3\%$.}
	\vspace{3mm}
	\label{tab2}
	\begin{tabular}{|c|c|c|c|}
		\hline
		($J_x/J_z$, $J_y/J_z$)&(0.45, 0.75)&(0.6, 0.75)&(0.75, 0.85)\\
		\hline
		$c$& 1.029 & 1.020 & 1.026 \\
		\hline
	\end{tabular}
\end{table}

\begin{figure}
	\includegraphics[width=0.45\textwidth]{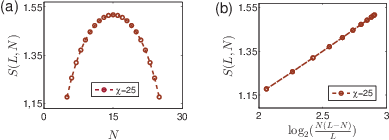}
	\caption{(color online) The entanglement entropy $S(L,N)$ for the chosen point $(J_x/J_z, J_y/J_z)=(0.2, 0.9)$ in the ${\rm CF}_{x}$ phase, with the system size $L$ being 30:
		(a) the entanglement entropy $S(L,N)$ versus the block size $N$; (b) the entanglement entropy $S(L,N)$ versus the finite-size universal scaling function $\log_2(N(L-N)/L)$.
		Here, $N$ ranges from 6 to 24, and the bond dimension $\chi$ is chosen to be $\chi=25$. 
	} 
	\label{SfiniteCF}
\end{figure}

We turn to the $\rm{CF_{x}}$ phase. Actually, the ground state $|\varphi_0\rangle$ on the characteristic line  $J_x/J_z=0$ with $J_y/J_z > 0$ is highly entangled, in addition to highly entangled degenerate ground states on the characteristic line $J_y/J_z=1$ with $J_x/J_z<1$. Hence,
one may anticipate that, even away from the two characteristic lines, the  entanglement entropy $S(L,N)$ for a ground state in the $\rm{CF_{x}}$ phase must scale in a similar way.
For a chosen point $(J_x/J_z, J_y/J_z)= (0.2, 0.9)$ in the ${\rm CF}_{x}$ phase, the model~(\ref{xyz2}) is numerically simulated to yield the ground state wave function in the MPS representation by means of the variational finite-size MPS algorithm~\cite{smfrank}.
In Fig.~\ref{SfiniteCF}, the entanglement entropy $S(L,N)$, as a function of $N$ for fixed $L$, is shown, with the system size $L=30$ and the bond dimension $\chi=25$. 
The entanglement entropy $S(L,N)$ does not saturate with increasing $N$ until $N$ reaches $L/2$, as expected. In addition, a finite-size scaling analysis is performed, with a universal finite-size scaling function $g(L,N)$ being $g(L,N)=N(1-N/L)$~\cite{finitesizesm}. As it turns out, the connection between the prefactor and the number of type-B GMs is lost.

\subsection{Emergent permutation symmetry  group $S_{L/2}$ in the ${\rm CF}_x$ phase}

In the ${\rm CF}_x$ phase,  an approximate permutation symmetry group $S_{L/2}$ emerges that becomes {\it exact} in the thermodynamic limit.

For our purpose, we have to evaluate the fidelity $F(L)$, which essentially compares the ground state $|\varphi_0\rangle$ with the state $P|\varphi_0\rangle$, resulted from the action of a permutation operation $P$ on $|\varphi_0\rangle$. Mathematically, the permutation group $S_{L/2}$ is generated from $L/2-1$ generators $P_{ll+1}$, representing the exchange operations acting on the two nearest-neighbor unit cells $l$ and $l+1$ ($l=1,\cdots,L/2-1$). As a result, we have $F_{ll+1}(L) = |\langle \varphi_0| P_{ll+1}|\varphi_0\rangle |$ ($l=1,\cdots,L/2-1$).

\begin {figure}
\includegraphics[width=0.48\textwidth]{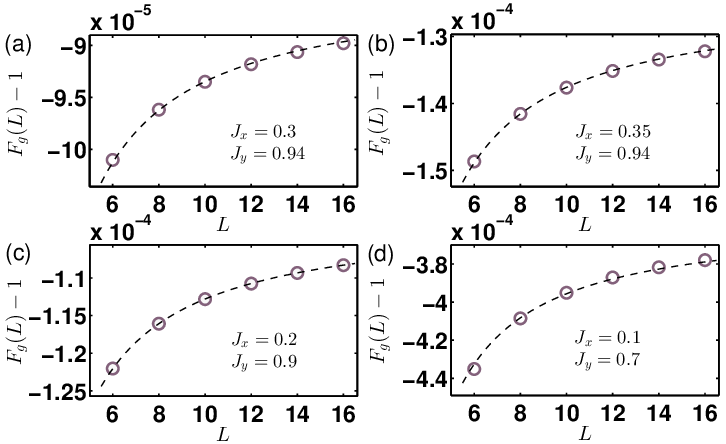}
\caption{(color online) The geometric average $F_g(L)$ of  $F_{ll+1}(L) = |\langle \varphi_0| P_{ll+1}|\varphi_0\rangle |$  ($l=1,\cdots,L/2-1$), as a function of the system size $L$ for the four chosen points: 
	(a) $(J_x/J_z, J_y/J_z)=(0.3, 0.94)$, (b) $(J_x/J_z, J_y/J_z)=(0.35, 0.94)$, (c) $(J_x/J_z, J_y/J_z)=(0.2, 0.9)$, 
	and (d) $(J_x/J_z, J_y/J_z)=(0.1, 0.7)$ in the ${\rm CF}_x$ phase, with $P_{ll+1}$ representing the exchange operations acting on the two nearest-neighbor unit cells $l$ and $l+1$ ($l=1,\cdots,L/2-1$), and the system size $L$ ranging from $6$ to $16$.}
\label{CFFid}
\end{figure}

\begin{table}
\renewcommand\arraystretch{1.5}
\caption{The parameter $R$ is extracted from the geometric average $F_g(L)=1- R(A \exp(\eta/L)/L+B\exp(-\kappa L)]$  for  four chosen points: (a) $(J_x/J_z, J_y/J_z)=(0.3, 0.94)$, (b) $(J_x/J_z, J_y/J_z)=(0.35, 0.94)$, (c) $(J_x/J_z, J_y/J_z)=(0.2, 0.9)$,
	and (d) $(J_x/J_z, J_y/J_z)=(0.1, 0.7)$
	in the ${\rm CF}_x$ phase, with the system size $L$ ranging from $6$ to $16$. Here, $A$, $B$, $\eta$ and $\kappa$ are taken from Table~\ref{tableEE},  extracted from the finite-size corrections to the ground state energy per lattice site, $e_0(L)$.}
\begin{tabular}{c|ccccccc}
	\hline
	\hline
	&$J_x/J_z=0.3$&$J_x/J_z=0.35$&$J_x/J_z=0.2$&$J_x/J_z=0.1$&\\
	&$J_y/J_z=0.94$&$J_y/J_z=0.94$&$J_y/J_z=0.9$&$J_y/J_z=0.7$&\\
	\hline
	\begin{minipage}{1.5cm} $R$ \end{minipage}
	& $0.885$ & $0.919$ & $0.875$ &$1.107$&\\
	\hline
	\hline
\end{tabular}
\label{tableFid}
\end{table}

Hence, we need to introduce the geometric average of $F_{ll+1}(L)$, denoted as  $F_g(L)$, as a measure to quantify the  extent to which the approximate  permutation symmetry $S_{L/2}$ for finite $L$'s deviates from the exact one:
\begin{equation}
F_g(L)=  \sqrt[L/2-1]{\prod_{l=1}^{L/2-1}{F_{ll+1}(L)}}.
\label{ga}
\end{equation}
The presence of the approximate  permutation symmetry $S_{L/2}$ implies that any two-point correlation function does not depend on the distance between the two points, up to $1/L^r$, with $r$ being a positive integer. Since this statement works for the Hamiltonian density in (\ref{xyz2}) and $P_{ll+1}$, one may conclude that the ground state energy per lattice site approaches $J_x^2$, in exactly the same way as $F_g(L)$ approaches 1. In other words, $e_0(L)-J_x^2$ should be proportional to $F_g(L)-1$. Equivalently, 
$F_g(L)$ scales as
\begin{equation}
F_g(L)=1- R(\frac{A e^{\eta/L}}{L}+B e^{-\kappa L}),\label{fidcf}
\end{equation}
where $R$ is a positive constant.

In Fig.~\ref{CFFid}, we plot $F_g(L)-1$ versus $L$ for four chosen points: (a) $(J_x/J_z, J_y/J_z)=(0.3, 0.94)$, (b) $(J_x/J_z, J_y/J_z)=(0.35, 0.94)$, (c) $(J_x/J_z, J_y/J_z)=(0.2, 0.9)$,
and (d) $(J_x/J_z, J_y/J_z)=(0.1, 0.7)$ in the ${\rm CF}_x$ phase, when $A$, $B$, $\eta$ and $\kappa$ are taken from  Table~.\ref{tableEE}, extracted from the finite-size corrections to the ground state energy  per lattice site, $e_0(L)$. 
The best fit is performed to yield the parameter $R$, as listed in Table~\ref{tableFid}.

\subsection{Finite-size corrections to the ground state energy $E_0$: the energy expectation value ${\bar E}$ for a H-orthogonal state}

On the characteristic line $J_x=0$ with $J_y/J_z>0$, a factorized ground state takes the form $|\phi_f(L)\rangle=\bigotimes_l |v_1v_2\rangle _l$~\cite{smspin1BM}, where $|v_1v_2\rangle _l = |v_1\rangle_{2l-1} |v_2\rangle _{2l}$,
with $|v_1 \rangle_{2l-1}$ and  $|v_2 \rangle_{2l}$ being a vector in a local spin space at the two nearest-neighbor lattice sites $2l-1$ and $2l$ ($l=1,\cdots,L/2$), respectively. They take the form
\begin{eqnarray}
	|v_1\rangle_{2l-1}=&\sin\zeta|0_y\rangle_{2l-1} + e^{i\theta}\cos\zeta |0_z\rangle_{2l-1},\\ \nonumber
	|v_2\rangle_{2l}= &\frac{J_y\cos\zeta}{\sqrt{J_y^2\cos^2\zeta+J_z^2\sin^2\zeta}}|0_y\rangle_{2l} +e^{-i\theta}\frac{J_z\sin\zeta}{\sqrt{J_y^2\cos^2\zeta+J_z^2\sin^2\zeta}}|0_z\rangle_{2l},
	\label{v12}
\end{eqnarray}
where $\zeta$ and $\theta$ are two free parameters that are real, and $|0_y\rangle_{2l-1/2l}$ and $|0_z\rangle_{2l-1/2l}$ are basis states, with an eigenvalue being zero, for the spin operators $S^y_{2l-1/2l}$ and $S^z_{2l-1/2l}$, respectively. Here, we  have introduced $L$ as an argument in a wave function to indicate the dependence on $L$.

We stress that  $|\phi_f(L)\rangle$ constitute a two-parameter family of ground states on the characteristic line $J_x=0$ with $J_y/J_z>0$. However, only $q=L+1$ states among them are linearly independent to each other. A convenient way to take advantage of this fact is to  exploit $V_q$ and $|\phi_0(L)\rangle$, already defined in the main text, to introduce $q$ H-orthogonal states  $|\psi_k\rangle$ on the characteristic line $J_x=0$ with $J_y/J_z>0$:  $|\psi_k(L)\rangle \equiv (V_q)^k|\phi_0(L)\rangle$~\cite{smshiqq2} ($k=0,1,\cdots,L$). Hence, it is the cyclic group ${\rm Z}_{q}$ that connects the $q$ H-orthogonal states  $|\psi_k(L)\rangle$, which becomes ${\rm U}(1)$ in the thermodynamic limit.
For convenience, the explicit expression for $|\phi_0(L)\rangle$ is cited:
$|\phi_0\rangle=\bigotimes_j |v\rangle_j$, where $|v \rangle_j=\mu |0_y\rangle_j + \nu |0_z\rangle_j$, with $\mu^2+\nu^2=1$ and  $\mu=\sqrt {J_y/(J_y+J_z)}$.

It is plausible to assume that, away from the characteristic line $J_x=0$ with $J_y/J_z>0$ in the $\rm{CF}_x$ phase, for a fixed value of $J_y$, the $q$ H-orthogonal states  $|\psi_k\rangle$ may be expanded into an asymptotic series, with a leading term being proportional to  $(V_q)^k |\phi_0(L)\rangle$. For our purpose, we focus on $|\psi_0(L)\rangle$, which takes the form
\begin{equation}
	|\psi_0(L)\rangle=\frac{1}{\sqrt{N_0}} \sum_n \omega_n |\phi(L,n)\rangle,
	\label{aa}
\end{equation}
where  $|\phi(L,n)\rangle$ denote a set of orthonormal states: $\langle\phi(L,m)|\phi(L,n)\rangle=\delta_{mn}$, and $\omega_n$ ($n=0,1,\cdots)$ denote the coefficients 
in $|\phi(L,n)\rangle$, with $N_0=\sum_j|\omega_j|^2$. Without loss of generalities, one may assume that $\omega_0$ is a positive number.
In particular, we set $|\phi(L,0)\rangle=|\phi_0(L)\rangle$, which is permutation-invariant with respect to the unit cells consisting of the two nearest-neighbor lattice sites .

Our aim is to determine $|\psi_0(L)\rangle$ up to the first order correction. This amounts to determining  $|\phi(L,1)\rangle$, which may be achieved if we act the Hamiltonian  $\mathscr{H}$, as presented in Eq.(\ref{xyz2}), on $|\phi(L,0)\rangle$ successively. As a result, we have
\begin{eqnarray*}
	\mathscr{H}|\phi(L,0)\rangle=J_x^2L|\phi(L,0)\rangle+a\sqrt{L}|\phi(L,1)\rangle,\\
	\mathscr{H}|\phi(L,1)\rangle\simeq a\sqrt{L}|\phi(L,0)\rangle  +  J_x^2 L(1-\frac{b}{L})|\phi(L,1)\rangle,
\end{eqnarray*}
where $a=J_x(J_z-J_y)$ and $b=3$. Note that $a$ is a positive number in the $\rm{LL}$ phases. Here, $|\phi(L,1)\rangle$, orthogonal to $|\phi(L,0)\rangle$, takes the form
\begin{equation}
	|\phi(L,1)\rangle =\frac{1}{\sqrt{C_{L/2}^1}}\sum_P|\underbrace{vv\cdots vv}_{L/2-1}\underbrace{|0_x0_x\rangle}_1,
\end{equation}
where the sum is taken over all the permutations $P$ for a given partition, with $L/2-1$ and $1$ denoting the numbers of the unit cells in  $|vv\rangle$ and $|0_x0_x\rangle$, respectively.
If we choose $\omega_1=u/\sqrt{L}$, with $u$ being a real number to be determined, then $N_{0}$  takes the form
\begin{equation}
	N_{0} \approx \sqrt {\omega_0^2+\omega_1^2} =\sqrt {\omega_0^2+|u|^2/L}.
\end{equation}

Now we are ready to evaluate the energy expectation value ${\bar E}(L)$ for a H-orthogonal state, which takes the form
\begin{equation}
	{\bar E}(L)=J_x^2L-\frac{2|u|a}{\omega_0}+O(\frac{1}{L}).
\end{equation}
Here, we have chosen $u$ to be negative to ensure that ${\bar E}(L)$ is less than $J_x^2L$. As follows from the Cauchy-Schwarz inequality: $|u|a\leq(|u|^2+a^2)/2$, we have
\begin{equation}
	{\bar E }(L)\geq J_x^2L-\frac{|u|^2+a^2}{\omega_0}+O(\frac{1}{L}), \label{correction}
\end{equation}
where the equality is valid if $|u|=a$.
In other words, it is necessary to choose $|u|=a$, in order to satisfy the physical requirement that the ground state energy  $E_0(L)$ must be as low as possible, given the relation between ${\bar E}(L)$ and $E_0(L)$ in Eq.(\ref{HO}). Hence,  we have
\begin{equation}
	{\bar E}(L)= J_x^2L-\frac{2 a^2}{\omega_0}+O(\frac{1}{L}).
\end{equation}

If we proceed to the next order in the asymptotic series in Eq.\;(\ref{aa}), it is possible to figure out the sub-leading correction $-g/L$, with $g$ being a constant. Instead, we restrict ourselves to pointing out that $g$ must be positive. Physically, this is due to the fact that a ground state yields the lowest energy expectation value. Indeed, if $g$ were negative, then it would yield a higher energy expectation value. If so, we should have stopped to proceed in the first place. In other words, the asymptotic series in Eq.\;(\ref{aa}) would terminate, but  obviously that is not the case. In fact, we attempt to approximate the ground state $|\varphi_0\rangle$ and the $L$ low-lying states $|\varphi_k\rangle$ ($k=1,\cdots, L$) in terms of $q$ permutation-invariant $H$-orthogonal states, as seen from Eq.(\ref{aa}). However, the permutation symmetry group $S_{L/2}$ is approximate for finite $L$'s, but becomes exact when $L$ tends to infinity. Hence, we have ${\bar E}(L)= J_x^2L-\frac{2 a^2}{\omega_0}-\frac{g}{L}$, with $g$ being positive, which may be rewritten as follows
\begin{equation}
	{\bar E}(L)\approx J_x^2L-A e^{\eta/L},
\end{equation}
where  $A=2a^2/\omega_0$ and $\eta= \omega_0 g/(2 a^2)$. Substituting into Eq.(\ref{HO}), we are led to the finite-size corrections to the ground state energy $E_0(L)$ in Eq.(\ref{el}).

Our construction above shows that the ground state and the low-lying states are very close to each other, which 
explains why it is challenging to simulate the spin-1 ferromagnetic anisotropic biquadratic model (\ref{xyz2})~\cite{smspin1BM}  in the ${\rm CF}_x$ phase by means of the tensor network algorithms~\cite{smTN2n2,smDM30}. Therefore, it becomes important to look into the model from an ED perspective.

A few remarks are in order. First, the second term in Eq.(\ref{correction}) originates from the emergent permutation symmetry in the ground state subspace and the third term in Eq.(\ref{correction}) originates from an alternative SSB pattern for $\rm{U(1)}$. Both of them vanish when the thermodynamic limit is approached. Second, it is the finite-size corrections to the ground state energy that mark an essential difference between the ${\rm CF}_x$ phase and  the $\rm{LL_{yz}}$ phase. The former is scale-invariant, but not conformally invariant, whereas the latter is conformally invariant, with central charge being one, subject to  the finite-size corrections to the ground state energy predicted from conformal field theory~\cite{smcardy,smaffleck}. 
Third, our asymptotic analysis suggests that the $q$-orthogonal states  $|\psi_k\rangle$, up to the first-order correction, are permutation-invariant, and
the finite-size corrections to the ground state energy takes the same form as that from a heuristic argument for generic permutation-invariant states.

\begin{figure}
	\includegraphics[width=0.44\textwidth]{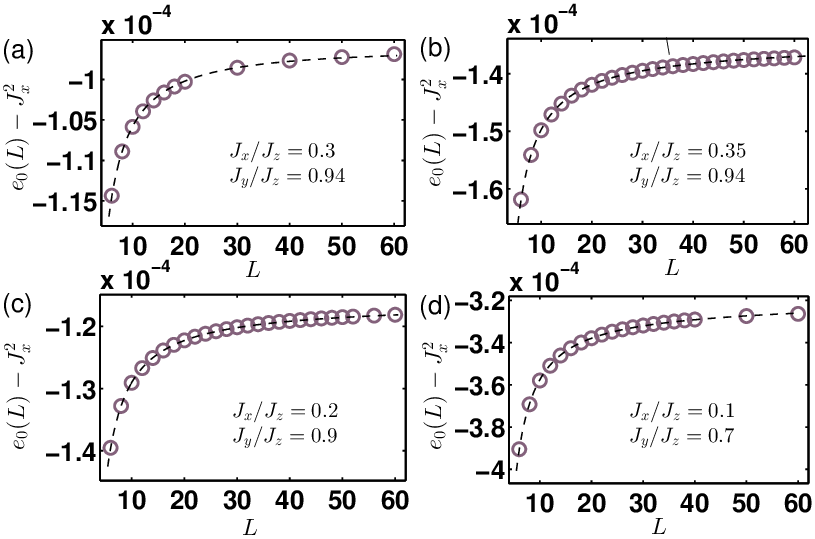}
	\caption{(color online) The finite-size corrections to the ground state energy per lattice site, denoted as $e_0(L)$, for four chosen points: (a) $(J_x/J_z, J_y/J_z)=(0.3, 0.94)$, (b) $(J_x/J_z, J_y/J_z)=(0.35, 0.94)$, (c) $(J_x/J_z, J_y/J_z)=(0.2, 0.9)$, and	(d) $(J_x/J_z, J_y/J_z)=(0.1, 0.7)$
		in the ${\rm CF}_x$ phase. Here, the finite-size DMRG algorithm is exploited to simulate the model (\ref{xyz2}) under PBCs, with the system size $L$ ranging from $6$ to $60$.}
	\label{EEFig}
\end{figure}

\subsection{Finite-size correlations to the ground state energy per lattice site $e_0(L)$ from the finite-size DMRG simulations}
\label{EnergyDMRGsm}

The ground state energy per lattice site, denoted as $e_0(L)$, is evaluated from the finite-size DMRG simulations~\cite{smDM10,smDM20} for the model Hamiltonian (\ref{xyz2})
under PBCs. The finite-size corrections to the ground state energy per lattice site, $e_0(L)$,  take the form
\begin{equation}
	e_0(L)={J_x}^2 -A \frac {e^{{\eta}/{L}}}{L} - B e^{-\kappa L}.
\label{energy}
\end{equation}
Indeed,  the second term originates from the emergent permutation symmetry in the ground state subspace,
which in turn is relevant to a gapped GM when the symmetry group $\rm{SU}(2)\times \rm{U}(1)$ on the
characteristic line $J_y = J_z$ is explicitly broken to $\rm{U}(1)\times \rm{U}(1)$, away from the characteristic line $J_y = J_z$. Meanwhile,
the third term originates from an alternative SSB pattern for $\rm{U(1)}$. The finite-size corrections to the ground state energy mark an essential difference between the ${\rm CF}$ phases and the $\rm{LL}$ phases. Here, we emphasize that the presence of $\exp(\eta/L)$ in the second term represents the emergence of a length scale, in addition to another length scale arising from ${\rm Z}_q$ in the third term. Therefore, two length scales are involved, competing with each other, in the ${\rm CF}_x$ phase.

In the main text, we have performed the best fit for the ground state energy per lattice site, $e_0(L)$
against the theoretical prediction in Eq.(\ref{energy}). Here, the best fit is performed for other four chosen points deep inside the ${\rm CF}_x$ phase.
In Fig.~\ref{EEFig}, we plot $e_0(L)$ versus $L$ for the four chosen points.
Our simulation results for $A$ and $B$, $\eta$ and $\kappa$ are listed in Table~\ref{tableEE}. As we have seen,
$A$ and $B$ vanish, as $J_x/J_z$ gets close to 0 and  $J_y/J_z$ gets close to 1, since no finite-size corrections arise on the two characteristic lines $J_x/J_z=0$ and  $J_y/J_z=1$. 

\begin{table}
	\renewcommand\arraystretch{1.5}
	\caption{The parameters $A$, $B$, $\eta$ and $\kappa$ are extracted from the finite-size corrections to the ground state energy per lattice site, denoted as $e_0(L)$, for the model (\ref{xyz2}), with the system size $L$ ranging from $6$ to $60$.}
	\begin{tabular}{c|ccccccc}
		\hline
		\hline
		&&$J_x/J_z=0.3$&$J_x/J_z=0.35$&$J_x/J_z=0.2$&$J_x/J_z=0.1$&\\
		&&$J_y/J_z=0.94$&$J_y/J_z=0.94$&$J_y/J_z=0.9$&$J_y/J_z=0.7$&\\
		\hline
		\begin{minipage}{1cm} $A$ \end{minipage}
		& & $0.919\times 10^{-4}$ & $1.221\times 10^{-4}$ & $1.079\times 10^{-4}$ &$2.735\times 10^{-4}$&\\
		\begin{minipage}{1cm} $B$ \end{minipage}
		& & $0.952\times 10^{-4}$ & $1.355\times 10^{-4}$ & $1.165\times 10^{-4}$ &$3.231\times 10^{-4}$&\\
		\begin{minipage}{1cm} $\eta$ \end{minipage}
		& & $1.4$ & $1.6$ & $1.5$ &$2.4$&\\
		\begin{minipage}{1cm} $\kappa$ \end{minipage}
		& & $0.580\times 10^{-4}$ &$0.645\times 10^{-4}$ & $0.303\times 10^{-4}$ & $0.976\times 10^{-4}$ &\\
		\hline
		\hline
	\end{tabular}
	\label{tableEE}
\end{table}

Note that the finite-size corrections to the ground state energy per lattice site, $e_0(L)$, are always negative, in the ${\rm CF}_x$ phase (also cf.~Ref.~\cite{smspin1BM}), thus implying that the ground state energy per lattice site, $e_0(L)$,  approaches $J_x^2$ from below, as $L$ tends to infinity.
Hence, as follows from the $q$ H-orthogonal states,  the symmetry group $\rm{U}(1)$, viewed
as a limit of $Z_q$ as $q\rightarrow\infty$, is spontaneously broken in the thermodynamic limit. As a consequence,  no gapless GM emerges, thus leading to  an alternative SSB pattern for $\rm{U(1)}$. The apparent contradiction with the Goldstone theorem~\cite{smGM0,smGM1,smGM2} requires clarification of the semantic meaning for continuous SSB.
That is, the dichotomy between continuous symmetry groups and discrete symmetry groups is not necessarily identical to that between continuous SSB and discrete SSB.

\subsection{Acknowledgements}

We are grateful to Murray Batchelor, John Fjaerestad and Ian McCulloch for comments and suggestions to improve the manuscript.

\end{document}